\documentclass[aps, pra, showpacs, amsmath, twocolumn]{revtex4}
\usepackage{graphicx}
\usepackage{dcolumn}
\usepackage{bm}
\usepackage{amsmath}

\begin{document}

\title{Intrinsic relation between ground-state fidelity and the
characterization of a quantum phase transition}
\author{Shu Chen, Li Wang, Yajiang Hao and Yupeng Wang}
\affiliation{Institute of Physics, Chinese Academy of Sciences, Beijing
100080, China}
\date{\today }

\begin{abstract}
The notion of fidelity in quantum information science has been
recently applied to analyze quantum phase transitions from the
viewpoint of the ground-state (GS) overlap for various many-body
systems. In this work, we unveil the intrinsic relation between
the GS fidelity and the derivatives of GS energy and find that
they play equivalent role in identifying the quantum phase
transition. The general connection between the two approaches
enables us to understand the different singularity and scaling
behaviors of fidelity exhibited in various systems on general
grounds. Our general conclusions are illustrated via several
quantum spin models which exhibit different kinds of QPTs.
\end{abstract}
\pacs{03.65.Ud, 64.60.-i, 05.70.Jk, 75.10.-b}
\maketitle

%05.70.Fh    phase transitions: general studies
%03.67.Mn    entanglement production, characterization, and manipulation
%03.75.Gg    entanglement and decoherence Bose-Einstein condensates
%03.75.Hh    static properties of condensates, thermodynamical statistical
%            and structural properties
%75.10.jm    quantized spin models
%75.30.kz    magnetic phase boundaries (including magnetic transitions,
%            metamagnetism, etc)
%03.67.-a    quantum information
%64.60.-i General studies of phase transitions (see also 63.70.+h
%Statistical mechanics of lattice vibrations and displacive phase transitions;
%for critical phenomena in solid surfaces and interfaces, and in
%magnetism, see 68.35.Rh, and 75.40.-s, respectively)
%75.10.-b General theory and models of magnetic ordering (see also
%05.50.+q Lattice theory and statistics)
%03.65.Ud Entanglement and quantum nonlocality (e.g. EPR paradox,
%Bell's inequalities, GHZ states,
%05.70.Jk Critical point phenomena in thermodynamics
%05.70.Fh, 03.67.Mn, 03.75.Gg, 03.75.Hh

% PACS Number

\section{Introduction}
Quantum phase transitions (QPTs) that happened at the zero
temperature is purely a phenomenon of ground state (GS) transition
driven by external parameters. Traditionally, QPTs are described
in terms of order parameter and symmetry breaking within the
Landau-Ginzburg paradigm which have been extensively studied in
condensed matter physics \cite{Sachdev}. In recent years, QPT has
also attracted a lot of attention in quantum-information science
\cite{Osterloh,Vidal}, in which one of the research focus is the
role of quantum entanglement in characterizing QPTs
\cite{Osterloh,Vidal,Wu,Gu,LAmico07}. More recently, another
concept in quantum information science, i.e., the fidelity has
been put forward to identify QPTs from the perspective of the GS
wave functions \cite {Zanardi06,HTQuan2006}. The GS fidelity is
defined as the overlap between two ground states with only
slightly different values of the external parameters
\cite{Zanardi06} and thus is a pure geometrical quantity. Since no
a priori knowledge of the order parameter is needed, the fidelity
might be a potential universal criteria for characterizing the
QPTs
\cite{Zanardi06,Pzanardi0606130,Buonsante,Chen,MFYang,HQZhou,HQZhou07,Gu07}.
An increasing interest has been drawn in the role of GS fidelity
in detecting QPTs for various many-body systems
\cite{Pzanardi0606130,Buonsante,Chen,MFYang,HQZhou,HQZhou07,Gu07},
since Zanardi and Paunkovic first exploited it to identify QPTs in
the XY spin chain \cite{Zanardi06} where the fidelity shows a
narrow drop at the transition point. Remarkably, the success of
fidelity analysis in dealing with the Bose-Hubbard model \cite
{Buonsante} and spin systems \cite{Chen,MFYang} implies that it
may have practical relevance even for more complicate strongly
interacting systems where no a simple description is possible. The
relation between the fidelity and Berry phase \cite{Zhu} has also
been unveiled in terms of Riemannian metric tensors
\cite{Zanardi07PRL}.

Generally one may expect that the GS fidelity shows an abrupt drop
in the vicinity of the QPT point of the system as a consequence of
the dramatic change of the structure of the GS. This is true for
the first-order QPTs caused by a level crossing of GSs due to the
GS wavefunctions at the different sides of the level-crossing
point are almost orthogonal (orthogonal at the level-crossing
point). However, this is neither an obvious nor a general
conclusion for a continuous QPT where no GS level crossing occurs
and the GS evolves ``adiabatically" in the parameter space
\cite{Gu07,Chen}. Conventionally, QPTs are characterized by
singularities of the ground state energy: first-order QPTs are
characterized by discontinuities in the first derivative of the
energy, whereas second-order ($n$th-order) QPTs are characterized
by discontinuities in the second ($n$th) derivative of the energy.
An important question is whether the singularity of GS energy is
intrinsically related to the GS overlap? Answering this question
is no doubt significant for a deeper understanding of QPTs, and
the validity and limitation of fidelity as a measure of QPTs.

So far the studies of fidelity as a measure of QPTs are based on
the analysis of particular many-body models
\cite{Zanardi06,Buonsante}. The general connection between GS
fidelity and QPTs is not yet well established. In this work, we
shall discuss, in a general framework, how the GS fidelity can be
related to a QPT characterized by non-analyticities of the GS
energy. Our result shows that the singularity and scaling
behaviors of the GS fidelity (or it derivatives) are directly
related to its correspondences of the derivative of GS energy.

Our paper is organized as follows. In section II, we reveal the
general relation between the fidelity susceptibility and the 2nd
order derivative of the GS energy. The subsequent sections are
devoted to two examples which exhibit the 2nd order and
Kosterlitz-Thouless (KT) QPTs, respectively. A summary is given in
the last section.

\section{Fidelity and quantum phase transition}
The general Hamiltonian of a quantum
many-body system undergoing QPTs reads
\begin{equation}
H(\lambda )=H_0+\lambda H_1,  \label{Ham}
\end{equation}
where $H_1$ is supposed to be the driving term with $\lambda $ the
control parameter. In terms of the eigenstates $|\Psi _n(\lambda
)\rangle $ of $H(\lambda )$, the Hamiltonian can be reformulated as $%
H(\lambda )=\sum_0^{\mathcal{N}-1}E_n(\lambda )|\Psi _n(\lambda
)\rangle \langle \Psi _n(\lambda )|,$ where $\mathcal{N}$ is
dimensions of the Hilbert space. The GS fidelity is defined as the
overlap between $ |\Psi _0(\lambda )\rangle $ and $|\Psi
_0(\lambda +\delta )\rangle $, i.e.
\begin{equation}
F(\lambda ,\delta )=\left| \langle \Psi _0(\lambda )|\Psi _0(\lambda +\delta
)\rangle \right| ,
\end{equation}
where $\Psi _0(\lambda )$ is the GS wavefunction corresponding to
the parameter $\lambda $ and $\delta $ is a small quantity
\cite{Zanardi06}. It is obvious that the fidelity is dependent of
$\delta $. The rate of change of fidelity is given by the second
derivative of fidelity\cite{Zanardi06} or fidelity susceptibility
(FS) \cite {Gu07}
\begin{equation}
S(\lambda )=\partial _\delta ^2F(\lambda ,\delta )\mid _{\delta =0}\simeq
2\lim_{\delta \rightarrow 0}\frac{1-F(\lambda ,\delta )}{\delta ^2},
\label{FS}
\end{equation}
which is $\delta $ independent and sometimes a more effective quantity to
detect the QPT.

First we discuss the first order QPT ($1$QPT) which is induced by
GS level crossing.
We assume $E_0(\lambda )<$ $E_1(\lambda )$ for $\lambda <\lambda _c$ with $%
\lambda _c$ being the crossing point where the GS energy level
$E_0$ crosses over the first excited level $E_1$, therefore we
must have $E_1(\lambda )<E_0(\lambda )$ for $\lambda >\lambda _c.$
Explicitly, for the $1$QPT, the GS energy is defined as
\[
E_g(\lambda )=\left\{
\begin{array}{lll}
E_0(\lambda ) &  & \lambda <\lambda _c, \\
E_1(\lambda ) &  & \lambda >\lambda _c.
\end{array}
\right.
\]
The derivative of the GS energy is defined as
\[
\frac{\partial E_g(\lambda )}{\partial \lambda }=\left\{
\begin{array}{lll}
\lim_{\lambda \rightarrow \lambda _c-0}\frac{E_0(\lambda )-E_0(\lambda _c)}{%
\lambda -\lambda _c}=\frac{\partial E_0\left( \lambda \right) }{\partial
\lambda } &  & \lambda <\lambda _c, \\
\lim_{\lambda \rightarrow \lambda _c+0}\frac{E_1(\lambda )-E_1(\lambda _c)}{%
\lambda -\lambda _c}=\frac{\partial E_1\left( \lambda \right) }{\partial
\lambda } &  & \lambda >\lambda _c.
\end{array}
\right.
\]
In general, $\frac \partial {\partial \lambda }E_0\left( \lambda
\right) $ is not equal to $\frac \partial {\partial \lambda
}E_1\left( \lambda \right) ,$ therefore the first derivative of GS
energy for a $1$QPT is not continuous at the transition point.
Accordingly, the GS fidelity at the transition point is defined as
$F(\lambda _c,\delta )=\left| \langle \Psi _g(\lambda _c-\delta
/2)|\Psi _g(\lambda _c+\delta /2)\rangle \right| =\left| \langle
\Psi _0(\lambda _c-\delta /2)|\Psi _1(\lambda _c+\delta /2)\rangle
\right| ,$ therefore we have $\lim_{\delta \rightarrow 0}F(\lambda
_c,\delta )=0$ which means that there appears a sharp drop in the
transition point. It is straightforward that the sudden drop in
the transition point shares the same physical origin with the
discontinuity of the first derivative of GS energy.

Next we consider the continuous phase transition for which the GS
of the Hamiltonian is nondegenerate for a finite system. For the
case of $2nd$ order QPT, there is no level crossing for the GS
energy, therefore we can always represent $E_g\left( \lambda
\right) =E_0\left( \lambda \right)$. The first derivative of the
GS energy is given by $\frac \partial {\partial \lambda }E_0\left(
\lambda \right) =\left\langle \Psi _0(\lambda )\right| H_1\left|
\Psi _0(\lambda )\right\rangle ,$ which is noting else but the
Hellmann-Feynman theorem. It is straightforward to get the second
derivative of GS energy $\frac{\partial ^2}{\partial \lambda^2
}E_0\left( \lambda \right) =\left\langle \Psi _0(\lambda )\right|
H_1\left| \partial _\lambda \Psi _0(\lambda )\right\rangle
+h.c.\text{ .}$ Inserting the identity operator
$\sum_{n=0}^{\mathcal{N}-1}\left| \Psi _n(\lambda )\right\rangle
\left\langle \Psi _n(\lambda )\right| =1$ between $H_1$ and $
\left| \partial _\lambda \Psi _0(\lambda )\right\rangle $, we then
get $ \frac{\partial ^2}{\partial \lambda^2 }E_0\left( \lambda
\right) =\sum_{n=1}^{\mathcal{N}-1}H_{0n}^1\left\langle \Psi
_n(\lambda )\right. \left| \partial _\lambda \Psi _0(\lambda
)\right\rangle +h.c.$ with $ H_{0n}^1=\left\langle \Psi _0(\lambda
)\right| H_1\left| \Psi _n(\lambda )\right\rangle .$ We note that
the term of $n=0$ is not included in the summation because it
gives zero due to $\left\langle \partial _\lambda \Psi _0(\lambda
)\right. \left| \Psi _0(\lambda )\right\rangle +\left\langle \Psi
_0(\lambda )\right. \left| \partial _\lambda \Psi _0(\lambda
)\right\rangle =0.$ Differentiating the eigenequation $H\left|
\Psi _0(\lambda
)\right\rangle =E_0\left| \Psi _0(\lambda )\right\rangle $ with respect to $%
\lambda $ and taking the inner product with $\Psi _n(\lambda )$ yields $%
\left\langle \Psi _n(\lambda )\right| \partial _\lambda H\left| \Psi
_0(\lambda )\right\rangle +\left\langle \Psi _n(\lambda )\right| H\left|
\partial _\lambda \Psi _0(\lambda )\right\rangle =\partial _\lambda
E_0\delta _{n0}+E_0\left\langle \Psi _n(\lambda )\right. \left| \partial
_\lambda \Psi _0(\lambda )\right\rangle .$ Exploiting the hermiticity of $H$
to write $\left\langle \Psi _n(\lambda )\right| H\left| \partial _\lambda
\Psi _0(\lambda )\right\rangle =E_n\left\langle \Psi _n(\lambda )\right.
\left| \partial _\lambda \Psi _0(\lambda )\right\rangle ,$ it follows that
for $n\neq 0,$ $\left\langle \Psi _n(\lambda )\right. \left| \partial
_\lambda \Psi _0(\lambda )\right\rangle ={\left\langle \Psi _n(\lambda
)\right| \partial _\lambda H\left| \Psi _0(\lambda )\right\rangle }/{%
(E_0-E_n)}.$ Therefore, we have
\begin{equation}
\frac{\partial ^2}{\partial \lambda ^2 }E_0\left( \lambda \right)
=\sum_{n\neq 0}\frac{2\left| \left\langle \Psi _n(\lambda )\right|
H_1\left| \Psi _0(\lambda )\right\rangle \right| ^2}{E_0(\lambda
)-E_n(\lambda )}. \label{2dE}
\end{equation}
To see the relation of FS with the second derivative of GS energy,
we rederive the expression of FS \cite {Gu07,Zanardi07PRL} by
expanding the wavefunction $\left| \Psi _0(\lambda +\delta
)\right\rangle $ in the basis of eigenstates corresponding to the
parameter $\lambda $, to the first order, which leads to
\[
\left| \Psi _0(\lambda +\delta )\right\rangle =c\left( \left| \Psi
_0(\lambda )\right\rangle +\delta \sum_{n\neq 0}\frac{H_{n0}^1\left( \lambda
\right) \left| \Psi _n(\lambda )\right\rangle }{E_0(\lambda )-E_n(\lambda )}%
\right) ,
\]
where $c=\sqrt{1+\delta ^2\sum_{n\neq 0}\left| H_{n0}^1\left( \lambda
\right) \right| ^2/[E_0(\lambda )-E_n(\lambda )]^2}$ is the normalization
constant. It follows directly $F(\lambda ,\delta )=c.$ Substituting it back
into Eq. (\ref{FS}), we get
\begin{equation}
S\left( \lambda \right) =\sum_{n\neq 0}\frac{\left| \left\langle \Psi
_n(\lambda )\right| H_1\left| \Psi _0(\lambda )\right\rangle \right| ^2}{%
\left[ E_0(\lambda )-E_n(\lambda )\right] ^2}.  \label{FSR}
\end{equation}

Comparing Eqs. (\ref{FSR}) and (\ref{2dE}), we find that for both
of them the singularities come from the vanishing energy gap in
the thermodynamic limit. However, a divergence does not
necessarily occur in the critical point, for example, for the KT
QPT as we shall discuss later, no singularity arises even the
energy gap tends to zero in the thermodynamic limit as matrix
elements $ H_{n0}^1$ may also vanish simultaneously.
%Now it is clear that the FS and second derivative
%of GS energy share same physical origin and thus play equivalent
%role in identifying QPT.
Despite the apparent similarity between Eqs. (\ref{FSR}) and
(\ref{2dE}), we note that fidelity susceptibility cannot be
expressed by the GS energy 2nd derivatives alone \cite{0801.2473}.
Nevertheless, the different power in the denominators in Eqs.
(\ref{FSR}) and (\ref{2dE}) shows that fidelity susceptibility
might be a more sensitive tool to detect critical points because
it can be more singular than the second derivative of GS energy.
Therefore, for the case where the GS energy 2nd derivative is
divergent at the critical point, it is no doubt that the fidelity
susceptibility is divergent too. However, for the case where the
second derivative of GS energy is not divergent, (for example the
3rd-order QPTs and the KT phase transition), it is hard to judge
whether the fidelity susceptibility becomes singular or not at
these critical points just from the relation between Eqs.
(\ref{FSR}) and (\ref{2dE}).

\section{Transverse field Ising model}
To understand the
equivalence of the two approaches in characterizing the quantum
criticality, we first apply them to study a specific example, say,
the transverse field Ising model with the Hamiltonian given by
\begin{equation}
H(\lambda )=-\sum_{i=1}^L\left( \hat{\sigma}_i^x\hat{\sigma}_{i+1}^x+\lambda
\hat{\sigma}_i^z\right)
\end{equation}
where $\lambda $ represents external magnetic field along the $z$
axis and the periodic boundary is assumed. Following the standard
procedure \cite{Sachdev,Lieb}, the model can be diagonalized by
applying Jordan-Wigner transformation which maps the spin model to
a fermion model and then using the Bogoliubov transformation
$c_k=b_k\cos \frac{\theta _k}2-ib_{-k}^{\dagger }\sin \frac{\theta
_k}2$. The result is given by $H=\sum_k\omega _k\left(
c_k^{\dagger }c_k-1\right) $ with $\omega _k=\sqrt{\sin ^2\frac{2\pi k}%
L+\left( \lambda -\cos \frac{2\pi k}L\right) ^2},$ where the summation is
over $k=-M,$ $\cdots ,$ $M$ with the assumption of $L=2M+1.$ The ground
state of $H$ is the vacuum $\left| g\right\rangle $ of $c_k$ given by $%
\left| g\right\rangle =\prod_{k=1}^M\left( \cos \frac{\theta
_k}2\left| 0\right\rangle _k\left| 0\right\rangle _{-k}-i\sin
\frac{\theta _k}2\left| 1\right\rangle _k\left| 1\right\rangle
_{-k}\right) ,$ with the ground state energy $E_0\left( \lambda
\right) =-\sum_k\omega _k$, where $\left| 0\right\rangle _k$ and
$\left| 1\right\rangle _k$ are, respectively, the vacuum and
single excitation of the $k$th mode $b_k$ and $\theta _k$ is
defined by $\sin \theta _k=\sin \frac{2\pi k}L/\omega _k.$ It
follows that the fidelity is given by $F(\lambda ,\delta
)=\prod_{k=1}^M\cos \frac{\theta _k-\widetilde{\theta }_k}2$ with
$\widetilde{\theta }_k=\theta _k\left( \lambda +\delta \right) $
\cite{Zanardi06}. After some very simple algebras, one get $
S(\lambda )=2\frac{1-F(\lambda ,\delta )}{\delta ^2}=\frac
14\sum_{k=1}^M\left( \frac{\partial \theta _k}{\partial \lambda
}\right) ^2. $ In the thermodynamic limit with $L\rightarrow
\infty $, the summation can be replaced by the integral and thus
we get
\[
\frac{\partial ^2}{\partial \lambda^2} e_0\left( \lambda \right)
=-\frac 1{2\pi }\int_{-\pi }^\pi \frac{\sin ^2\varphi }{\left[
\sin ^2\varphi +(\lambda -\cos \varphi )^2\right] ^{3/2}} d\varphi
\]
and
\[
s(\lambda )=\frac 1{16\pi }\int_{-\pi }^\pi \frac{\sin ^2\varphi
}{\left[ \sin ^2\varphi +(\lambda -\cos \varphi )^2\right] ^2}
d\varphi
\]
where $e_0\left( \lambda \right) =E_0\left( \lambda \right) /L$
and $s(\lambda )=S(\lambda )/L$. Both of the above integrals are
divergent when $\lambda =1$. Their singular behavior can be
analyzed in the vicinity
of critical point $\lambda _c=1$ with the asymptotic behavior described by $%
-\partial ^2e_0\left( \lambda \right) /\partial \lambda
^2=-0.31187\ln |\lambda -\lambda _c|+const$ and $\ln s(\lambda
)=-\ln |\lambda -\lambda _c|+const$, respectively.
\begin{figure}[tbp]
\includegraphics[width=3.3in]{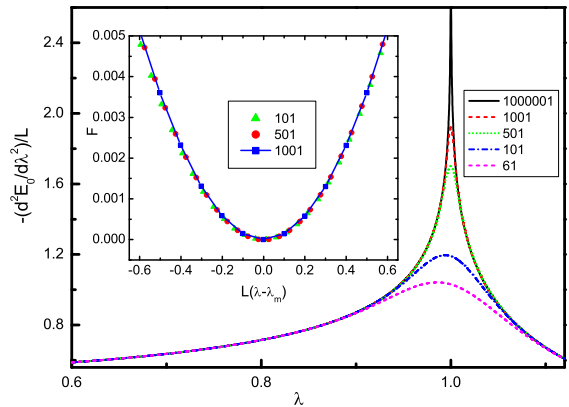}
\caption{(color online) The derivatives $\partial ^2E_0\left( \lambda
\right) /\partial \lambda ^2$ as a function of $\lambda $.}
\label{fig1}
\end{figure}

To get further insight on how the two approaches work, we study
both the scaling behaviors of FS and second derivative of GS
energy. As shown in Fig. 1, we display the second derivatives of
GS energy for different lattice sizes. Despite no real divergence
for finite $L$, it is obvious that the curves exhibit marked
anomalies with height of peak increasing with the lattice size.
The parameter $\lambda _m$ labelling the position of peak
approaches the critical point $\lambda _c=1$ in a way of
$\lambda_m =1- const L^{-1.802}$. The value of peak diverges
logarithmically with increasing lattice size as $ -\partial
^2e_0(\lambda ,L)/\partial \lambda ^2|_{\lambda _m}=0.31132\ln {L}
+const$. By proper scaling scheme, we can fit all the data of
$F=\left[ 1-\exp (\partial ^2e_0/\partial \lambda ^2|_{\lambda
_m}-\partial ^2e_0/\partial \lambda ^2)\right] $ as a function of
$L\left( \lambda -\lambda _m\right) $ for different $L$ into a
single curve as displayed in the inset of Fig. 1. Correspondingly,
the FS exhibits similar scaling behaviors as shown in Fig. 2.
Around the critical point, all the data for different $L$ collapse
into a single curve of $F=\left[ 1-s(\lambda)/s(\lambda_m) \right]
$ as a function of $L\left( \lambda -\lambda _m\right) $. A
related quantity has been used in \cite{HQZhou} to make scaling
analysis for the quantum Ising model. It is
clear that $\partial ^2e_0\left( \lambda \right) /\partial \lambda ^2$ and $%
s(\lambda )$ exhibit similar critical behavior around the critical
point.
\begin{figure}[tbp]
\includegraphics[width=3.3in]{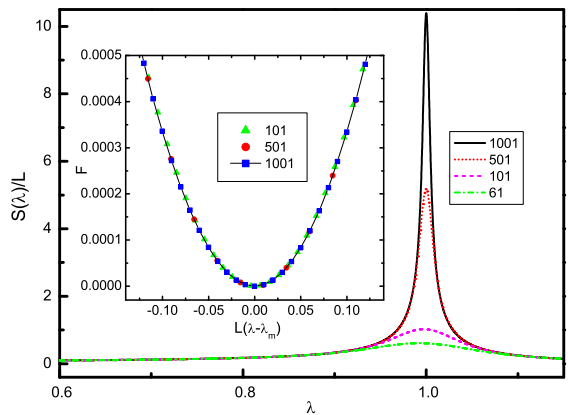}
\caption{(color online) The fidelity susceptibility as a function
of $ \lambda $.} \label{fig2}
\end{figure}

\section{$XXZ$ spin chain}
Having demonstrated the
equivalent role of the two different approaches in identifying QPT
on the transverse Ising model, we now use this procedure to
understand the more elusive KT-type QPT for which, contrary to the
previous expectation, the GS fidelity and the FS turn out to be
insensitive to QPT in some cases, for example, in the spin
$J_1-J_2$ model \cite{Chen}. We now consider the $XXZ$ model which
provides a further test for the intrinsic relationship between the
derivatives of GS energy and FS. The model is defined by
\begin{equation}
H(\lambda )=-\frac{1}{2}\sum_{i=1}^L\left(
\hat{\sigma}_i^x\hat{\sigma}_{i+1}^x
+\hat{\sigma}_i^y\hat{\sigma}_{i+1}^y +\lambda
\hat{\sigma}_i^z\hat{\sigma}_{i+1}^z\right)
\end{equation}
with $\lambda $ the exchange anisotropy parameter and it is related to $%
-H(-\lambda )$ by a unitary transformation. It is well known that
the model is in a critical phase for $-1\leq \lambda <1$, an
antiferromagnetic phase at $\lambda <-1$, and ferromagnetic phase
at $\lambda >1.$ The QPT at the ferromagnetic isotropic point
$\lambda =1$ is a 1QPT caused by the level crossing of GS. Here we
shall focus on the QPT at the antiferromagnetic isotropic point
$\lambda =-1$ which is known to be of KT type. In the whole
parameter regime the model is integrable. The eigenstate of
$H(\lambda )$ has the form of $\left| \Psi \right\rangle
=\sum_{n_1<\cdots <n_M}a(n_1,\cdots ,n_M)S_{n_1}^{-}\cdots
S_{n_M}^{-}\left| F\right\rangle $ with the coefficients having
the Bethe-ansatz form \cite{MTakahashib}
\[
a(n_1,\cdots ,n_M)=\sum_P\exp [i\sum_\alpha k_{p_\alpha }n_\alpha +\frac
i2\sum_{\alpha <\beta }\theta (k_{p_\alpha },k_{p_\beta })]
\]
where $P$ is any permutation of $(n_1,\cdots ,n_M)$ and the phase
$\theta (p,q)=2\tan ^{-1}\left( \frac{\lambda \sin [(p-q)/2]}{\cos
[(p+q)/2]-\lambda \cos [(p-q)/2]}\right) $. The eigenenergies
$E=-\frac \lambda 2 L +2 \sum_i (\lambda -\cos k_i )$ are
determined by a set of quasi-momentum $p_i$ ($ i=1,\cdots ,M)$ for
$M$ down spins which are the solution of Bethe ansatz equations
(BAEs) $ Nk_i=2\pi I_i-\sum_{j=1}^M\theta (k_i,k_j) $, where $I_i$
are integer or half-odd integer.
\begin{figure}[tbp]
\includegraphics[width=3.6in]{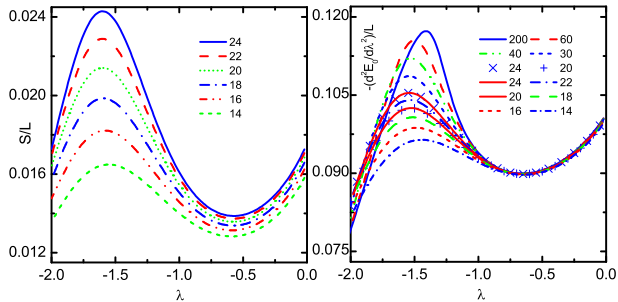}
\caption{(color online) The fidelity susceptibility and second
order derivative of GS energy for $XXZ$ model as a function of $
\lambda $. The data labelled by $\times$ and $+$ are obtained by
ED and they comply with the data by Bethe-ansatz method exactly.}
\label{fig3}
\end{figure}

By numerically solving the BAEs, we can get the GS state energy
for different sizes of $L$. As shown in Fig.3, we calculate the
$2$nd derivative of GS energy by both Bethe-ansatz and exact
diagonaliztion (ED) methods and the FS by ED. Our numerical
results show that there is no any singularity around the critical
point $\lambda _c=-1$ for both the FS and $2$nd order derivative
of GS energy for different sizes. For the GS energy derivatives,
this is true even in the thermodynamic limit where actually, as
proven in \cite{Yang}, all the $n$th order derivatives of GS
energy are continuous. Thus, in view of its connection to the
derivatives of GS energy, we can understand why the GS fidelity
failed to reproduce quantum critical behavior by the desired
singularities in the critical point for the KT transition. It is
also natural to explain why the finite size scaling of FS has a
clear distinction between noncritical and critical phases because
the GS energies are known to fulfil quite different finite size
scaling in critical or noncritical phase \cite{Hamer,Blote}. The
GS energy in the critical regime fulfills the finite-size
correction $ E_0(\lambda ,L)/L=e_0(\lambda )-\pi \upsilon \left(
\lambda \right) c/6L^2+o(L^{-2}) $ with $c=1$ being the conformal
anomaly number and $\upsilon =\pi \sin \gamma /\gamma $ the
``sound velocity'' for the model \cite{Hamer}. Here $\gamma$ is
defined by $ \lambda=-\cos \gamma$. On the other hand, the
conformal invariance is broken for $\lambda<-1$ and the GS energy
fulfills a different finite size scaling \cite{Blote}. Bearing in
mind that the equivalent role of FS and derivative of GS energy,
we expect that the FS exhibits similar scaling behavior in the
critical regime as its counterpart of derivative of GS energy does
and there is a deviation from the scaling behavior when
$\lambda<-1$. Indeed, the FS has been shown to scale differently
in critical and noncritical phases based on the results by ED
method \cite{Zanardi07PRL}. This feature can be used as an
indicator of KT QPT occurring in this model. Before ending the
discussion, we would like to give a remark to the problem whether
the fidelity approach is suitable to detect the KT transition? As
no rigorous result of the FS even for the exactly solvable model
is available in the thermodynamic limit, therefore this question
is still in a doubtful status \cite{MFYang,0801.2473} and remains
an open problem.

\section{Conclusions}
We have shown the intrinsic relationship of
the fidelity and derivatives of GS energy and revealed their
equivalent role in identifying the QPTs. Within our framework, the
singularities and scaling behavior of fidelity for various systems
can be well understood on general grounds. Through concrete
examples, we display that divergence of the FS or derivatives of
GS energy can be applied to identify the 2nd order QPT, whereas
for the KT transition the criticality is not a sufficient
condition to ensure divergence of the FS or derivatives of GS
energy.

\begin{acknowledgments}
The authors are grateful to M. F. Yang for useful discussions by
correspondence and the referee for bringing our attention to Ref.
\cite{0801.2473}. This work is supported by NSF of China under
Grant No. 10574150, MOST grant 2006CB921300 and programs of CAS.
\end{acknowledgments}

\end{document}